\documentclass[
nofootinbib,
twocolumn,
english,
superscriptaddress,
floatfix,
amsmath,amssymb,
longbibliography
]{revtex4-2}
\usepackage{tikz}
\usetikzlibrary{circuits.ee.IEC}
\usepackage{graphicx}
\usepackage{dcolumn}
\usepackage{soul}
\usepackage{bm}
\usepackage[hidelinks]{hyperref}
\usepackage{siunitx}
\usepackage[normalem]{ulem}
\hypersetup{colorlinks = true, allcolors = blue}

\makeatletter
\def\Dated@name{}
\def\Dated@text{}
\def\@date{}
\def\print@date{}
\makeatother

\graphicspath{{Figs}}

\begin{document}

\righthyphenmin=4
\lefthyphenmin=4

\title{
Voltage dynamics of spherical membranes from single ion channel currents
}

\author{Sirui Ning}
\thanks{These authors contributed equally.}
\affiliation{University of California, Berkeley, CA 94720, USA\looseness=-1}

\author{Joshua B. Fernandes}
\thanks{These authors contributed equally.}
\affiliation{University of California, Berkeley, CA 94720, USA\looseness=-1}

\author{Karthik Shekhar}
\thanks{Correspondence: kranthi@berkeley.edu, kshekhar@berkeley.edu}
\affiliation{University of California, Berkeley, CA 94720, USA\looseness=-1}
\affiliation{Lawrence Berkeley National Laboratory, Berkeley, CA 94720, USA}

\author{Kranthi K. Mandadapu}
\thanks{Correspondence: kranthi@berkeley.edu, kshekhar@berkeley.edu}
\affiliation{University of California, Berkeley, CA 94720, USA\looseness=-1}
\affiliation{Lawrence Berkeley National Laboratory, Berkeley, CA 94720, USA}

\date{}

\begin{abstract}
Ion channels and pumps drive ion-selective currents through cell membranes at localized sites, yet a cell's electrical state is routinely summarized by a single transmembrane voltage. Combining theory and numerical simulations, we resolve the spatiotemporal dynamics of charge reorganization driven by a localized current on a spherical membrane vesicle. At early times, the response is insensitive to membrane geometry: as in the case of a flat membrane~\cite{row2025spatiotemporal,fernandes2026electrochemical}, the transmembrane voltage decays in a monopolar fashion, varying inversely with distance from the source, and crosses over to a dipolar tail that scales as the inverse cube of distance. Under sustained current, this monopolar response spreads outward from the source. Because the vesicle is closed, this response cannot persist indefinitely; once the monopolar front traverses the entire vesicle, the subsequent charging dynamics is dominated by a spatially uniform mode corresponding to capacitive charging of the membrane. 
We further decompose the bulk potentials into an electrostatic image-charge component that generates the bulk electric fields and a spatially uniform capacitive mode {that can be represented as} an equivalent circuit.
We also derive a nonlocal cable equation governing the transmembrane voltage dynamics and show that the uniform mode is its long-time solution. 
This work provides a first-principles basis for the electrophysiological simplification of an \emph{electrotonically compact cell}.
\end{abstract}

\maketitle

\section{Introduction}
In recent works~\cite{row2025spatiotemporal,fernandes2026electrochemical}, we investigated how charge reorganizes after a localized ionic current passes through a single channel in a flat membrane. Real cellular membranes, however, are not necessarily flat and may close into vesicles, wrap into tubes, or form arbitrarily curved cellular boundaries~\cite{sahu2020geometry}. This raises a natural question: \emph{how does membrane geometry influence the ionic reorganization dynamics generated by a localized transmembrane current?} 
The answer is directly relevant to classical electrophysiological idealizations, in which membrane patches, axons, or compact cells are described by transmembrane voltages coupled to capacitive and {conductive} currents~\cite{hodgkin1946electrical,hodgkin1952quantitative,rall1962theory}. 
Furthermore, for compact cells, it is often assumed that a single transmembrane potential describes the entire electrical state of the membrane; this is the electrotonically compact approximation \cite{rall1969time,brown1981passive,rall1992matching}. However, it remains unclear over what length scales and timescales this approximation becomes valid under a localized current.
In this work, we address these questions in a spherical membrane, which provides the simplest closed geometry for studying charge reorganization dynamics arising from localized ion-selective currents. 

\begin{figure}[t!]
    \centering
    \includegraphics[width=0.9\linewidth]{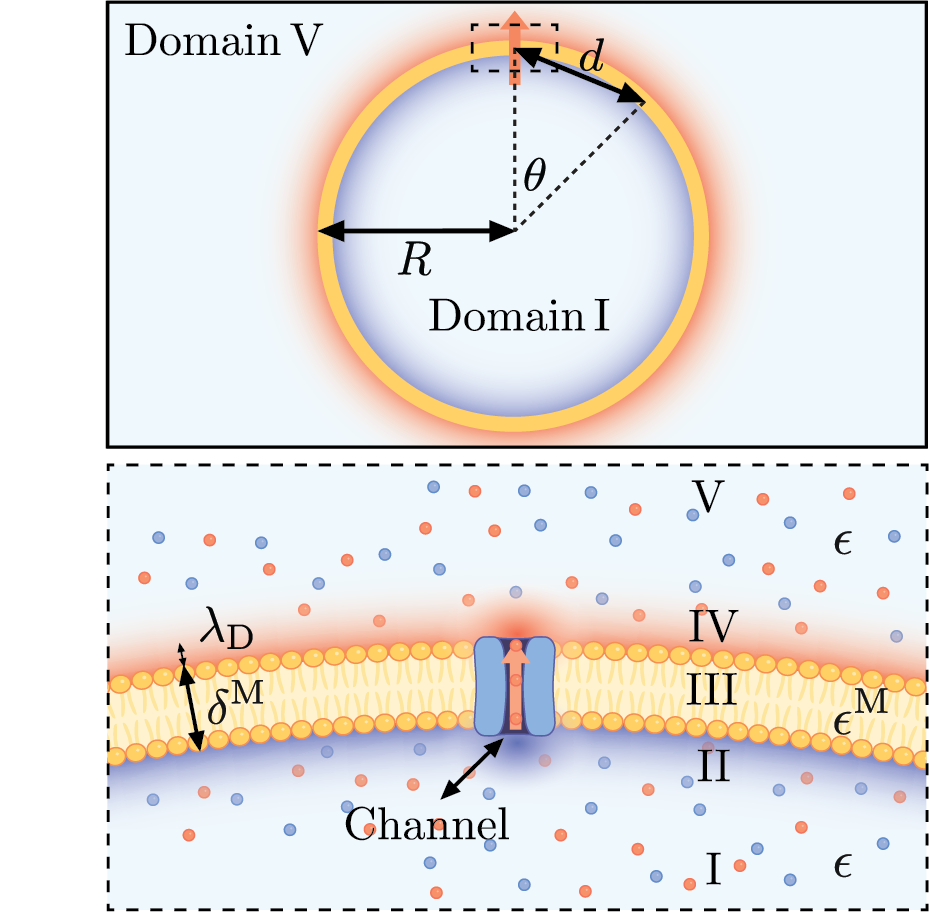}
    \caption{\textbf{Problem schematic.} A spherical lipid membrane of radius $R$, thickness $\delta^{\rm M}$, and permittivity ${\epsilon^{\rm M}\sim 4\epsilon_0}$ is immersed in a binary electrolyte solution of permittivity ${\epsilon \sim 80 \epsilon_0}$, where ${\epsilon_0}$ is the vacuum permittivity. The dielectric mismatch is ${\Gamma \equiv \epsilon / \epsilon^{\rm M} \sim 20}$. Starting at ${t=0}$, a cationic transmembrane current flows through a channel located at ${\theta=0}$. The resulting free charge is confined primarily to diffuse layers of characteristic thickness $\lambda_{\rm D}$ adjacent to the two membrane surfaces, as shown in the inset. The inner electrolyte, inner diffuse charge layer, membrane, outer diffuse charge layer, and outer electrolyte are denoted by domains I--V, respectively.}
    \label{fig:schematic}
\end{figure} 

We begin by briefly summarizing the results of Refs.~\cite{row2025spatiotemporal,fernandes2026electrochemical} for flat membranes. In an initially electroneutral background, a localized cationic current creates a charge imbalance in the diffuse charge (Debye) layers of thickness ${\lambda_{\rm D} \sim 1 \text{ nm}}$ adjacent to the membrane, generating long-ranged electric fields that drive rapid in-plane charge reorganization. At very early times, comparable to the Debye timescale ${\tau_{\rm D} = \lambda_{\rm D}^2/D \sim 1 \text{ ns}}$, where $D$ is the ion diffusivity, the transmembrane potential $V^{\rm M}(r,t)$ develops a monopolar region ${V^{\rm M}\sim 1/r}$ for ${r\lesssim \Gamma\delta^{\rm M}}$, surrounded by a dipolar region ${V^{\rm M}\sim 1/r^3}$,  where $r$ is the in-plane distance from the channel. Here, ${\Gamma \sim 20}$ is the bulk-to-membrane dielectric mismatch, and ${\delta^{\rm M}\sim 4}$~nm is the membrane thickness. Such power law profiles for ${V^{\rm M}}$ reflect the long-ranged character of the electrostatic field produced when an ion-selective transmembrane current breaks electroneutrality across the membrane. Under a sustained current, the front separating the monopolar and dipolar regions expands at constant velocity ${v = g/(2\mathcal{C_{\rm eq}})}$, where ${g \equiv \epsilon D/\lambda_{\rm D}^2 }$ is the bulk ionic conductivity, {$\epsilon$ is the bulk permittivity} and $\mathcal{C_{\mathrm{eq}}}$ is the equivalent capacitance per unit area of the membrane and the two adjacent diffuse charge layers. Defining the diffuse-layer and membrane capacitances as ${\mathcal{C_{\rm D}}\equiv \epsilon/\lambda_{\rm D}}$ and ${\mathcal{C_{\rm M}} \equiv \epsilon^{\rm M}/\delta^{\rm M}}$, respectively, the equivalent capacitance is given by $1/\mathcal{C_{\rm eq}} = 2\lambda_{\rm D}/\epsilon + \delta^{\rm M}/\epsilon^{\rm M}$. 
In physiological conditions ($\sim 150$~mM 1:1 electrolyte), this gives ${v\sim 40}$~m/s, much faster than diffusion over cellular length scales.

A spherical vesicle, unlike an infinitely extended flat membrane, is closed and encloses a finite volume. For a vesicle of radius ${R\gg\lambda_{\rm D}}$, one may expect the early-time diffuse-layer response to behave as if it were \emph{locally flat} near the current source. However, this flat-membrane description must eventually break down as the electrical disturbance reaches the vesicle scale. Since the flat-membrane charging front propagates with speed ${v=g/(2\mathcal C_{\rm eq})}$, this breakdown is expected after a time ${\tau_{\rm C}\sim 2R/v}$,\footnote{{This estimate is analogous to the capacitive charging time of a vesicle in an electrolyte under an applied external electric field. In an RC circuit-like scaling, the timescale is ${\mathcal{R}_{\rm eq}\mathcal C_{\rm eq}}$~\cite{farhadi2025capacitive,fernandes2026electrochemical}, where ${\mathcal{R}_{\rm eq}= 2R/g}$ is the effective electrolyte resistance per unit area across a length scale of  $2R$.}} corresponding to tens of nanoseconds for a micron-sized vesicle under physiological conditions. At later times, the closed geometry constrains the electric field lines and requires charge redistribution over the entire membrane surface. The nature of this later-time electrical response, however, remains unclear. One may then ask: \emph{how does membrane closure shape the late-time dynamics? Moreover, when does the electrotonically compact approximation, in which a single voltage describes the entire membrane, become valid?}

To resolve the closed-vesicle charging dynamics, we analyze the Poisson--Nernst--Planck (PNP) equations on the sphere, building on the boundary-layer framework established in Ref.~\cite{fernandes2026electrochemical}. We consider the small-parameter limit ${\eta \equiv \lambda_{\rm D}/(2R) \ll 1}$, in which nonzero charge density is confined to thin diffuse charge layers {adjacent to the membrane}, while the bulk remains electroneutral. A perturbative expansion of the PNP equations in $\eta$ yields a reduced problem in which harmonic bulk {electrostatic} potentials are coupled to dynamically charging interfacial layers. The transmembrane potential first reproduces the flat-membrane monopolar--dipolar front propagating from the channel at speed $v$. When this front reaches the opposite pole at ${t\sim\tau_{\rm C}}$, however, a second front emerges, separating the monopolar region from a nearly uniform background and propagating back toward the source on the timescale ${\tau_{\rm EC}\sim \tau_{\rm C}R/\lambda_{\rm D} \gg \tau_{\rm C}}$. For ${t>\tau_{\rm EC}}$, the sphere behaves as an electrotonically compact object with an {approximately} uniform transmembrane potential whose magnitude increases in time.

Lastly, we show that {the transmembrane potential} $V^{\rm M}$ obeys a \emph{cable equation}~\cite{hodgkin1946electrical, rall1962theory, dayan2005theoretical}, in which membrane patches are coupled not by local currents along the membrane, but through the surrounding electrolyte; the resulting lateral conductance is therefore \emph{nonlocal}. In this closed geometry, all spatial variations in voltage decay on the timescale $\tau_{\rm C}$, while the spatially uniform component remains and carries the global charging.  The classical \emph{local} cable equation is recovered only when the external electrolyte is confined to a thin layer with a nearby boundary.

\section{Problem Formulation}\label{Sec2}
\noindent\textbf{Setup and governing equations.} We consider a spherical lipid membrane vesicle of mid-surface radius ${R\gg \lambda_{\rm D}}$ and thickness $\delta^{\rm M}$ separating identical binary monovalent electrolytes of initial concentration $C_0$ (Fig.~\ref{fig:schematic}). At ${t=0}$, a localized cationic current $I$, representing an ion channel or pump, is initiated at the north pole. Anticipating that charge separation is localized near the membrane while the bulk electrolytes remain largely electroneutral, we divide each electrolyte into a bulk domain and an adjacent diffuse charge layer. This subdivision results in five domains: the inner bulk electrolyte, inner diffuse charge layer, membrane, outer diffuse charge layer, and outer bulk electrolyte, denoted by domains I--V, respectively; see Fig.~\ref{fig:schematic}.  We use spherical coordinates $(r,\theta,\varphi)$, with the origin at the center of the vesicle and the polar axis chosen to pass through the source. The inner and outer bulk electrolyte domains occupy the regions ${0<r<R-\delta^{\rm M}/2}$ and ${r>R+\delta^{\rm M}/2}$, respectively, while the membrane occupies the region ${R-\delta^{\rm M}/2<r<R+\delta^{\rm M}/2}$. Ion concentrations within the membrane are neglected because the membrane is treated as an ion-impermeable dielectric. We seek the electrolyte concentrations $C_\pm^\alpha(\mathbf{x},t)$ and the electrostatic potential $\phi^\alpha(\mathbf{x},t)$ in each domain, with ${\alpha\in\{{\rm I},\,{\rm II},\,{\rm III},\,{\rm IV},\,{\rm V}\}}$. The choice of the polar axis makes the problem axisymmetric, so all fields depend only on $r$, $\theta$, and $t$, but not on the azimuthal angle $\varphi$; see the Supplemental Material (SM)~\cite{supplemental_material} for full details.

In the electrolyte domains, the concentrations evolve according to the PNP equations, which arise in the dilute limit of Onsager transport theory~\cite{Nernst1888,Nernst1889,Planck1890,fong2020transport}. They are given by
\begin{align}
    \frac{\partial C_\pm^\alpha}{\partial t} &= - \boldsymbol{\nabla}\cdot \mathbf{j}^\alpha_\pm \ , \label{eq:mass-balance}\\
    \mathbf{j}_\pm^\alpha &= -D\left(\boldsymbol{\nabla} C_\pm^\alpha + \frac{z_\pm{\rm e}}{k_{\rm B}T} C_\pm^\alpha \boldsymbol{\nabla}\phi^\alpha \right) \label{eq:mass-flux}\ ,
\end{align}
where $D$ is the ionic diffusivity, assumed equal for anions and cations, $k_{\rm B}T/{\rm e}$ is the thermal voltage, ${z_\pm=\pm 1}$ are the ion valences, and ${\alpha\in\{{\rm I},\,{\rm II},\,{\rm IV},\,{\rm V}\}}$. The electric potential follows Poisson's equation
\begin{equation}
    -\epsilon \nabla^2\phi^\alpha = \rho^\alpha \equiv {\rm e}(C_+^\alpha-C_-^\alpha) \label{eq:gauss}\ ,
\end{equation}
where $\rho^\alpha$ is the free charge density. {Since the membrane contains zero} free charge,  the potential inside satisfies Laplace's equation, ${\nabla^2\phi^{\rm III}=0}$.

The boundary conditions encode two physical constraints: anions are blocked at all interfaces, while cations cross the membrane only at the localized source. Therefore, ${\mathbf{j}_-^\alpha\cdot\mathbf{n}^\alpha=0}$, and the imposed cationic flux is
\begin{align}
    \mathbf{j}_+^{\rm II}\cdot\mathbf{n}^{\rm II} &= \frac{I}{\rm e}\frac{\delta(\theta)}{2\pi r^2\sin(\theta)}
    & &\text{at } r = R-\delta^{\rm M}/2 \ ,\label{eq:bc1} \\
    \mathbf{j}_+^{\rm IV}\cdot\mathbf{n}^{\rm IV} &= -\frac{I}{\rm e}\frac{\delta(\theta)}{2\pi r^2\sin(\theta)}
    & &\text{at } r = R + \delta^{\rm M}/2\label{eq:bc2} \ .
\end{align}
 Here, ${\mathbf{n}^{\rm II} = \boldsymbol{e_r}}$ and ${\mathbf{n}^{\rm IV} = -\boldsymbol{e_r}}$ are the outward-pointing normals from the inner and the outer membrane surfaces. The opposite signs in Eqs.~\eqref{eq:bc1}--\eqref{eq:bc2} represent the same cationic current leaving one electrolyte and entering the other. 
 Electrostatic boundary conditions require continuity of the potential and of the normal dielectric displacement across each interface, with permittivity $\epsilon$ in the electrolyte and $\epsilon^{\rm M}$ in the membrane.
 We neglect fixed membrane surface charge; if present, it would appear as a prescribed jump in the normal displacement.

$\\$\textbf{Boundary-layer theory.}~Under physiological conditions, ${\eta \equiv \lambda_{\rm D}/(2R) \ll 1}$, ${\delta^{\rm M}/(2R) \ll 1}$, and ${{\rm e}V^{\rm M}/k_{\rm B}T \sim 1}$. The last condition corresponds to transmembrane voltages on the order of the thermal voltage, ${V^{\rm M}\sim 25}$~mV, and requires only one excess ion per ${\sim\!10^3}$ lipid molecules, so bulk concentrations remain close to equilibrium and the response stays linear~\cite{fernandes2026electrochemical}. Consequently, the problem {has a clear separation of length scales.} Free charge is confined to nanometer-scale diffuse-charge layers {near the membrane}, while the electric field {varies over the much larger scale of} the vesicle. {This separation motivates a} boundary-layer analysis {in which} the diffuse layers {act} as capacitive  {regions} that dynamically accumulate charge while remaining coupled to the electrostatic fields {in the bulk}. In what follows, we summarize the bulk and boundary-layer {equations} and their associated matching conditions; see Ref.~\cite{fernandes2026electrochemical} and the SM for detailed derivations.

\textit{Bulk solutions.}~At any order in $\eta$, the bulk electrolytes maintain electroneutrality. Although this removes any free charge in the bulk, it does not eliminate the electric field. 
The injection of a localized current drives vesicle-scale ionic reorganization by adjusting the harmonic bulk electrostatic potentials satisfying
\begin{equation}\label{eq:laplace}
    \nabla^2 \phi^\alpha = 0\ , \qquad \alpha = {\rm I},\,{\rm V}\ .
\end{equation}
Thus, all free charge is stored in the interfacial layers, while the bulk acts as a conducting pathway that redistributes charge around the closed geometry.

\begin{figure}[t!]
    \centering
    \includegraphics[width=0.9\linewidth]{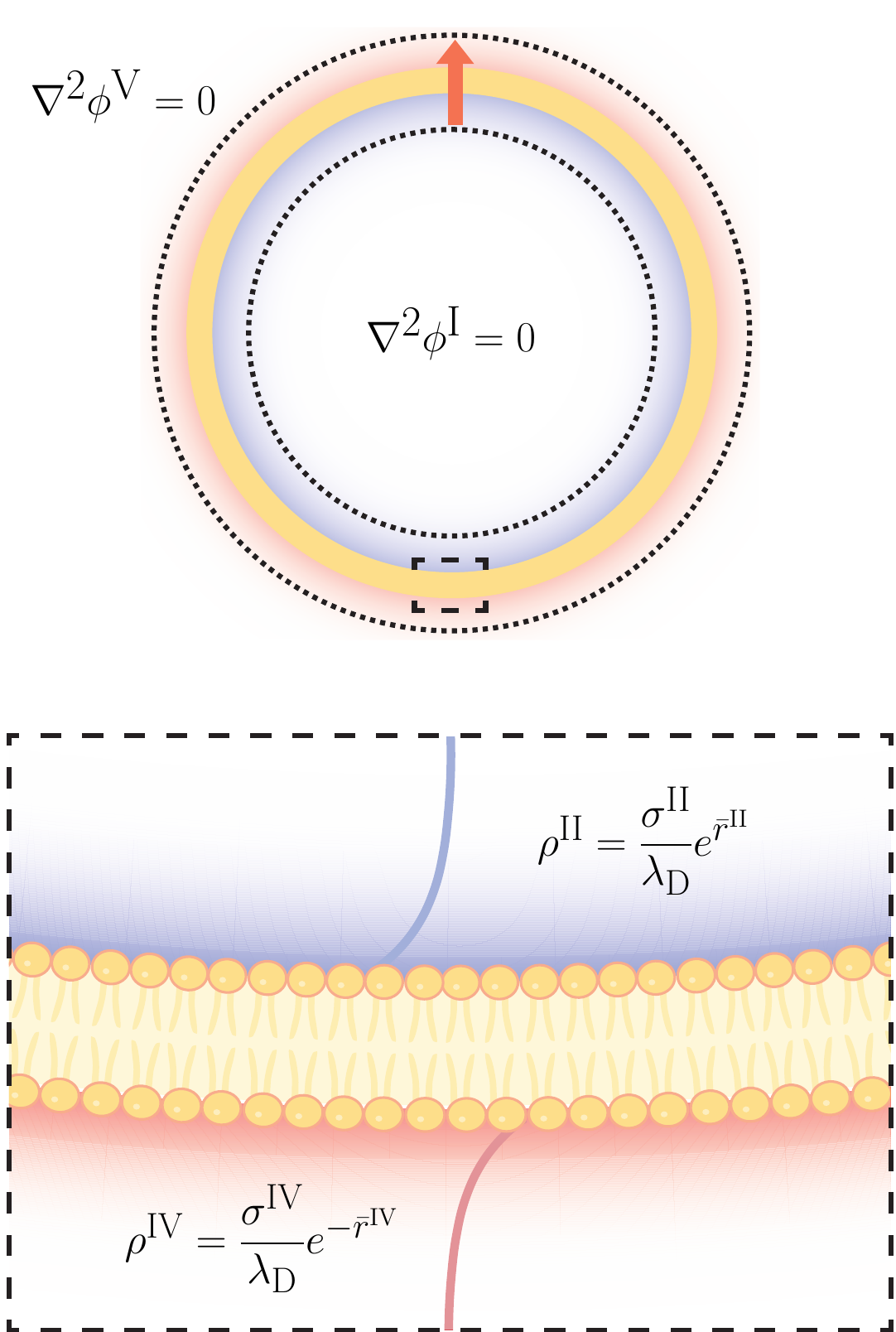}
    \caption{\textbf{Boundary-layer reduction.} For ${\eta \ll 1}$, the electrolyte bulk remains electroneutral and is described by harmonic potentials {obeying Laplace's equation}, while free charge is confined to diffuse layers of thickness ${\sim\lambda_{\rm D}}$ adjacent to the membrane. Within each diffuse layer, the charge density decays exponentially in the stretched radial coordinate. Each layer is represented in the reduced theory by an areal charge density $\sigma^\alpha(\theta,t)$ coupled to the bulk conduction problem.}
    \label{fig:boundary-layer}
\end{figure}

\textit{Diffuse charge layers.}~On timescales ${t\gg\tau_{\rm D}}$, the diffuse layers II and IV remain quasistatic to leading order in $\eta$. Furthermore, any spatial variations along the tangential direction on the sphere are of higher order compared to variations in the normal direction. Consequently, the leading-order problem becomes one-dimensional in the radial direction. In terms of stretched {radial} coordinates, for example 
${\bar r^{\rm IV} = (r-R-\tfrac{\delta^{\rm M}}{2})/\lambda_{\rm D}}$, 
the charge density obeys the linearized Poisson--Boltzmann equation and decays exponentially away from the membrane, ${\rho^\alpha = (\sigma^\alpha/\lambda_{\rm D}) e^{-|\bar r^\alpha|}}$ for ${\alpha \in \{{\rm II},{\rm IV}\}}$. Here, $\sigma^\alpha(\theta,t)$ is the areal charge density stored capacitively in diffuse layer $\alpha$, which sets the voltage drop across the membrane. The boundary-layer reduction for the spherical geometry is summarized in Fig.~\ref{fig:boundary-layer}.

\textit{Matching conditions at $r=R$.}~{The problem can be closed by} conditions {at the interface}. \emph{First}, conservation of charge in the outer diffuse layer gives
\begin{equation}\label{eq:charge-balance}
    \frac{\partial \sigma^{\rm IV}}{\partial t} = g\left.\frac{\partial \phi^{\rm V}}{\partial r}\right|_{r=R} + I\,\frac{\delta(\theta)}{2\pi R^2 \sin\theta}\ , 
\end{equation}
where the left side is the rate at which the diffuse layer accumulates charge, the first term on the right is the normal conduction current from the outer bulk electrolyte, and the second term is the imposed transmembrane current.
The inner diffuse layer obeys a corresponding balance with the opposite sign
\begin{equation}\label{eq:charge-balance2}
\frac{\partial \sigma^{\rm II}}{\partial t}
=-g\left.\frac{\partial \phi^{\rm I}}{\partial r}\right|_{r=R}-I\frac{\delta(\theta)}{2\pi R^2\sin\theta} \ .
\end{equation}
Equations~\eqref{eq:charge-balance}--\eqref{eq:charge-balance2} {denote} current conservation {in either boundary layer}: injected {current} is balanced by the rate of capacitive {charging of} the interfacial layers and the conductive current through the electroneutral bulk.

\emph{Second}, because the membrane contains no free charge, the electric potential satisfies Laplace's equation and varies linearly across the membrane thickness to leading order. Together with continuity of the normal dielectric displacement across the membrane and diffuse layers, this implies equal and opposite charging of the diffuse layers, ${\sigma^{\rm II}=-\sigma^{\rm IV}}$. Adding Eqs.~\eqref{eq:charge-balance} and~\eqref{eq:charge-balance2} then gives equality of the normal bulk potential gradients on the two sides of the membrane, i.e.,
\begin{equation}\label{eq:disp-cont}
    \left.\frac{\partial \phi^{\rm V}}{\partial r}\right|_{r = R}
    = \left.\frac{\partial \phi^{\rm I}}{\partial r}\right|_{r = R}\ .
\end{equation} 

\emph{Third}, {we match} the inner, outer, and membrane potentials {at the two membrane surfaces} and {use} ${\sigma^{\rm II}=-\sigma^{\rm IV}}$. {This} relates the transmembrane voltage ${V^{\rm M} = \sigma^{\rm IV}/\mathcal{C}_{\rm M}}$ to the potential {difference} between the two bulk solutions 
\begin{equation}\label{eq:VM-bulk}
    V^{\rm M}(\theta,t) = \frac{\chi}{1+\chi}\left(\left.\phi^{\rm V}\right|_{r=R} - \left.\phi^{\rm I}\right|_{r=R}\right)\ .
\end{equation}
Here, ${\chi \equiv \mathcal{C}_{\rm D}/2\mathcal{C}_{\rm M}}$ is the ratio of the diffuse-layer capacitance to the membrane capacitance. For physiological parameters ${\chi\sim 40 \gg 1}$~\cite{fernandes2026electrochemical}, so ${V^{\rm M}\approx \left.(\phi^{\rm V}-\phi^{\rm I})\right|_{r=R}}$. This means that most of the potential drop between the two bulk solutions appears across the low-permittivity membrane. The diffuse charge layers nevertheless remain essential because they store the compensating charge and screen the electroneutral bulk. Equations~\eqref{eq:laplace}--\eqref{eq:VM-bulk} close the boundary-layer problem.

\section{Results}\label{Sec3}
To solve the reduced boundary-layer problem, we expand the bulk potentials in Legendre modes {and} solve Eq.~\eqref{eq:laplace} {for each mode}. {We then} use the  diffuse {charge} layer balances (Eqs.~\eqref{eq:charge-balance} and \eqref{eq:charge-balance2}), together with Eq.~\eqref{eq:disp-cont} and potential-matching conditions above, to determine {how each mode evolves}. The localized current at the north pole fixes the Legendre coefficients of the source term, {and the resulting modes can be resummed to give} a closed-form expression for $V^{\rm M}$ {for all times ${t\gg \tau_{\rm D}}$}. {We compare this result with} numerical solutions of the fully nonlinear PNP equations using the finite element method (FEM), implemented in \texttt{FEniCSx} \cite{Scroggs22a,Scroggs22b,Alnaes14}. The {analytical and numerical} details are provided in the SM. {We first present} the closed-form expression for $V^{\rm M}$ {and} its short- and long-time limits, which reveal the crossover from the locally flat monopolar--dipolar response to a global, nearly uniform charging mode. {Next,} we analyze the bulk potentials, decomposing them into electrostatic and capacitive contributions. We then discuss the {electrochemical} relaxation {following} channel closure. {Finally, we show that} the voltage dynamics {can be cast as} an equivalent nonlocal cable equation.

\begin{figure}[t!]
    \centering
    \includegraphics[width=\linewidth]{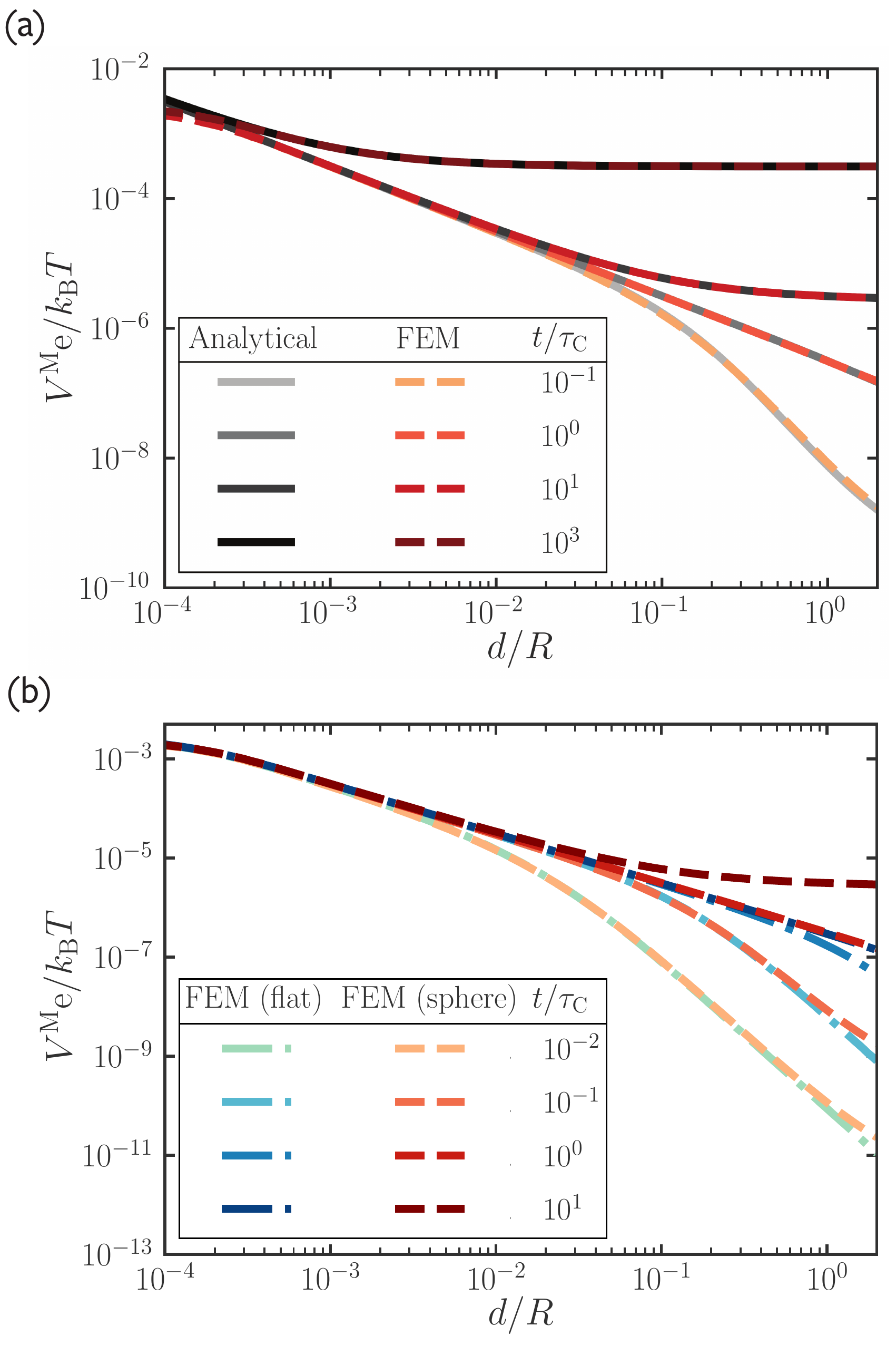}
    \caption{\textbf{Transmembrane potential along the vesicle surface.}
$V^{\rm M}$ as a function of the dimensionless chordal distance
${d/2R=\sin(\theta/2)}$ from the localized current source.
(a) The closed-form boundary-layer solution, Eq.~\eqref{eq:VM-main}
(solid curves), agrees quantitatively with FEM simulations of the full
nonlinear PNP equations (dashed lines) over four decades in ${t/\tau_{\rm C}}$, except very near the source, where the boundary-layer approximation breaks down. At early times, the profile exhibits the monopolar--dipolar structure of the
flat-membrane response~\cite{row2025spatiotemporal, fernandes2026electrochemical};
at late times, it becomes nearly uniform over the sphere, with only a residual
near-source contribution.
(b) Comparison between FEM simulations for the spherical vesicle and
for the flat membrane geometry considered in Refs.~\cite{fernandes2026electrochemical, row2025spatiotemporal}. At early times, the two profiles collapse, confirming
that the flat solution is the tangent-plane limit of the spherical theory.
They separate only when the charging front reaches distances
comparable to the vesicle radius, where the closed surface supports a global
capacitive charging mode.
Parameters: ${\Gamma=20}$, ${\delta^{\rm M}=4\lambda_{\rm D}}$,
${R=10^4\lambda_{\rm D}}$, ${I=2\times10^{-2}\,{\rm e}C_0D\lambda_{\rm D}}$,
and source radius ${R^{\rm P}=10^{-1}\lambda_{\rm D}}$. }
    \label{fig:results}
\end{figure}

$\\$\textbf{Closed-form transmembrane potential.}~The reduced problem yields a closed-form expression for $V^{\rm M}$ in terms of the dimensionless chordal distance $\hat d$ and time $\hat t$:
\begin{equation}\label{eq:VM-main}
\begin{split}
V^{\rm M}(\hat d,\hat t)
&=
\mathcal{V}_0
\Bigg[
\hat{t} + \frac{1}{2}
-2\left(1-\frac{e^{-\hat t}}{2}\right)^2
\\
&\qquad
+\frac{1}{\hat d}-\frac{1}{\sqrt{\hat d^2+\sinh^2\hat t}}
\\
&\qquad
-\log\!\left(
\frac{1+\hat d}
{\cosh\hat t+\sqrt{\hat d^2+\sinh^2\hat t}}
\right)
\Bigg]\ .
\end{split}
\end{equation}
Here, ${\mathcal{V}_0 = \frac{\chi}{1+\chi}\frac{I}{2\pi g R}}$, ${\hat t=t/\tau_{\rm C}}$, and ${\hat d=d/2R}$. The chordal distance from the source is ${d=2R\sin(\theta/2)}$, with ${d=0}$ at the channel and ${d=2R}$ at the opposite pole (Fig.~\ref{fig:schematic}). Equation~\eqref{eq:VM-main} is one of the central analytical results of the paper, derived in Sec.~S2 of the SM.
Figure~\ref{fig:results}(a) compares Eq.~\eqref{eq:VM-main} with FEM simulations of the full nonlinear PNP equations at representative times spanning four decades in $t$. The two agree quantitatively over the full range of ${d/R}$, except very near the source, confirming that the boundary-layer approach captures the essential physics of the full electrochemical problem. Near the source, on length scales ${d\lesssim\lambda_{\rm D}}$, the boundary-layer framework breaks down, and the solution is expected to cross over to the near-field regime identified in Ref.~\cite{row2025spatiotemporal}.

The dependence on ${\hat t=t/\tau_{\rm C}}$ in Eq.~\eqref{eq:VM-main} shows explicitly that the relevant timescale is neither the Debye relaxation timescale $\tau_{\rm D}$ nor the diffusion time across the vesicle, ${\tau_{\rm R} = (2R)^2/D}$, but {the capacitive timescale} ${\tau_{\rm C}}$, which is the time for the monopolar--dipolar charging front to traverse the vesicle. 
Equation~\eqref{eq:VM-main} {therefore has} two asymptotic regimes, ${t\ll \tau_{\rm C}}$ and ${t\gg \tau_{\rm C}}$, which we analyze below.

$\\$\textbf{Short-time dynamics and the recovery of the flat-membrane response.}~
In the limit ${t\ll \tau_{\rm C}}$, expanding Eq.~\eqref{eq:VM-main} at fixed $d$ and retaining the leading-order terms yields the transmembrane voltage
\begin{equation}\label{eq:VM-short-dim}
V^{\rm M}(d,t)
\simeq
\frac{\chi}{1+\chi}\,
\frac{I}{\pi g}
\left(
\frac{1}{d}
-
\frac{1}{\sqrt{d^2+(vt)^2}}
\right)\ .
\end{equation}
This is precisely the flat-membrane result, with the planar radial distance replaced by the chordal distance $d$~\cite{row2025spatiotemporal, fernandes2026electrochemical} (cf.~Eq.~(27) in Ref.~\cite{row2025spatiotemporal}). It describes a monopolar region near the source, where ${V^{\rm M}\sim 1/d}$, smoothly matched across ${d^*\sim vt}$ to a dipolar tail, where ${V^{\rm M}\sim (vt)^2/d^3}$, with the monopolar--dipolar front moving at speed $v$. At these times, when ${d^*\sim vt\ll R}$, the charging dynamics on the spherical surface is essentially indistinguishable from that of a flat membrane. {Indeed}, Fig.~\ref{fig:results}(b) demonstrates that the numerical solutions for spherical vesicles and flat membranes coincide at early times ${t\ll \tau_{\rm C}}$. Once the monopolar--dipolar front has propagated over a distance comparable to the vesicle radius, ${vt\sim R}$, corresponding to ${t\sim\tau_{\rm C}}$ or ${\hat t\sim1}$, the flat-membrane description breaks down and the effect of spherical geometry becomes important.

$\\$\textbf{Long-time dynamics: emergence of a uniform regime.}~
We next consider times ${t\gg \tau_{\rm C}}$, after the monopolar--dipolar front has traversed the scale of the vesicle. To determine the leading-order behavior in this regime, we expand Eq.~\eqref{eq:VM-main} in powers of $e^{-\hat t}$, obtaining
\begin{equation}\label{eq:VM-long-dim}
\begin{split}
V^{\rm M}(d,t)
& \simeq
\frac{\chi}{1+\chi}
\frac{I}{\pi g R}
\Bigg[
\frac{t}{\tau_{\rm C}} \\
&\qquad \qquad
+\frac{R}{d}
-\frac{3}{4}
-\frac{1}{2}\log\left(1+\frac{d}{2R}\right)
\Bigg]\ .
\end{split}
\end{equation}
Three features stand out. \emph{First}, the leading time-dependent term is spatially uniform, i.e., independent of $d$, and grows linearly in time under sustained current injection. Therefore, once the monopolar--dipolar front has traversed the entire surface, the closed membrane behaves increasingly as a global capacitor that is charged uniformly by the imposed current. \emph{Second}, the spatially dependent terms form a time-independent correction that dominates near the source through the residual monopolar contribution ${\sim R/d}$.
\emph{Third}, at a distance $d$ from the channel, the voltage becomes effectively uniform {when} the time-dependent term in Eq.~\eqref{eq:VM-long-dim} exceeds the leading monopolar correction.  Balancing these terms gives the crossover timescale ${t \sim {R\tau_{\rm C}}/{d}}$, or equivalently ${d\sim R\tau_{\rm C}/t}$. Thus, for ${t\gtrsim\tau_{\rm C}}$, the uniform region first appears near the antipode, where (${d=2R}$), and then expands back toward the current source as time increases. 
In this regime, the vesicle {becomes} increasingly electrotonically compact, so that a single transmembrane {voltage} describes {nearly} the entire vesicle. We define the vesicle as electrotonically compact {when} the boundary between the uniform and monopolar-dominated regions propagates back to within a Debye length of the source, ${d\sim \lambda_{\rm D}}$. {This gives} a timescale ${\tau_{\rm EC} \sim {R\tau_{\rm C}}/{\lambda_{\rm D}}}$, {which is about} ${50\,\mu{\rm s}}$ {for a micron-sized vesicle}.

\begin{figure}[t!]
    \centering
    \includegraphics[width=0.8\linewidth]{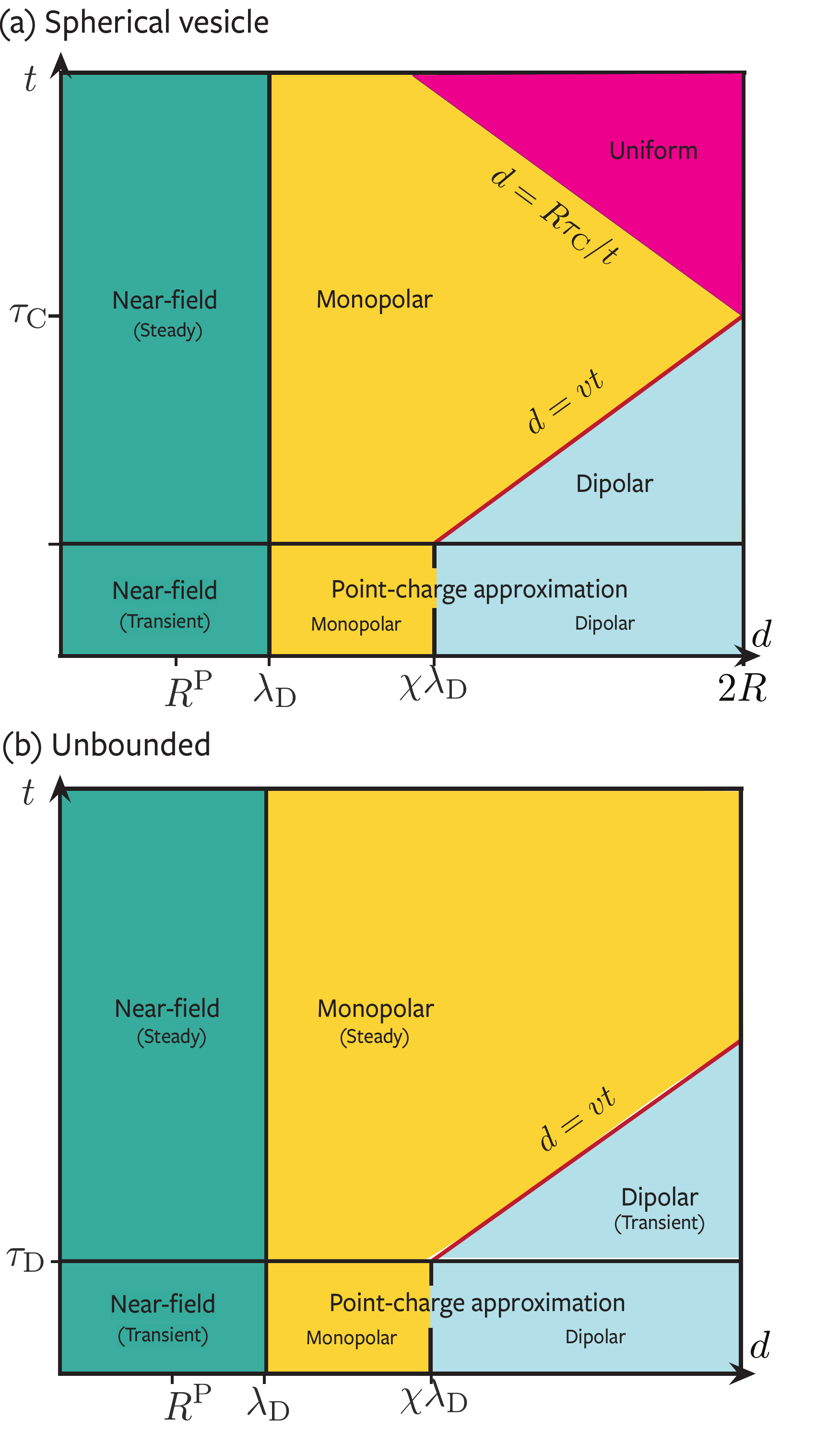}
    \caption{\textbf{Spacetime regimes on the spherical (top) and flat (bottom) membrane.} The closed spherical vesicle (top) develops a near-field region set by the pore radius $R^{\rm P}$, an outward-propagating monopolar--dipolar front for ${t\lesssim\tau_{\rm C}}$, and an inward-propagating monopolar--uniform front for ${t\gtrsim\tau_{\rm C}}$. The unbounded flat membrane (bottom) has no global length scale and therefore supports only the outward monopolar--dipolar crossover, with no late-time uniform regime. See Ref.~\cite{fernandes2026electrochemical} for a comparison with the flat membrane response in the presence of constant potential electrodes.}
    \label{fig:spacetime}
\end{figure}

$\\$\textbf{Spacetime structure of the transmembrane response.}~Combining the short- and long-time analyses gives a compact spacetime picture of the transmembrane-voltage response, summarized in Fig.~\ref{fig:spacetime}(a). The $(d,t)$ plane {is partitioned by two fronts}. The monopolar--dipolar front, ${d \sim vt}$, expands from the source at speed $v$ and reaches the {antipode} at ${t \sim \tau_{\rm C}}$, where the spherical geometry {terminates} the dipolar region. A second front, at ${d \sim R\,\tau_{\rm C}/t}$, separates the monopolar core from the uniform plateau. It appears first near the antipode and {moves} back toward the source as $1/t$. Thus, ${t\sim\tau_{\rm C}}$ marks the local-to-global crossover: local flatness controls the early response, while {the finite spherical geometry controls} the late response. By contrast, the flat-membrane response (Fig.~\ref{fig:spacetime}(b)) has no {uniform regime and contains only} the outward-propagating monopolar--dipolar front. The uniform plateau is therefore a direct consequence of closure, {since} a finite-area membrane supports a spatially uniform capacitive charging mode, whereas an infinite plane does not.

$\\$\textbf{Electric potentials and fields in the bulk.}
{We also} obtain analytical expressions for the electric potential {in} the inner and outer electrolytes. {We decompose these} bulk potentials {into electrostatic and capacitive contributions, which illuminate} the physical origin of the electric fields, the spatiotemporal propagation of the emergent electrical signal, and how the transmembrane potential is generated. The exact analytical expressions and derivations are provided in {Secs.~S2.3 and S2.4 of the SM}. {The} decomposition takes the form
\begin{align}
    \label{eq:phi_I_bulk}
    \phi^{\rm I}(r,\theta,t) &= \phi_{\rm el}^{\rm I}(r,\theta,t) - \phi_{\rm el}^{\rm I}(0,\theta,t) + \phi_{\rm cap}^{\rm I}(t) \ , \\
    \label{eq:phi_V_bulk}
    \phi^{\rm V}(r,\theta,t) &= \phi_{\rm el}^{\rm V}(r,\theta,t) - \lim_{r\to\infty}\phi_{\rm el}^{\rm V}(r,\theta,t) \ ,
\end{align}
where $\phi_{\rm el}^\alpha$ is the electrostatic potential {generated by} a set of image charges. The capacitive term $\phi_{\rm cap}^{\rm I}(t)$ encodes the potential difference between the interior potential at ${r=0}$ and the exterior potential as ${r\to\infty}$. 

\begin{figure*}[t!]
    \centering
    \includegraphics[width=0.8\linewidth]{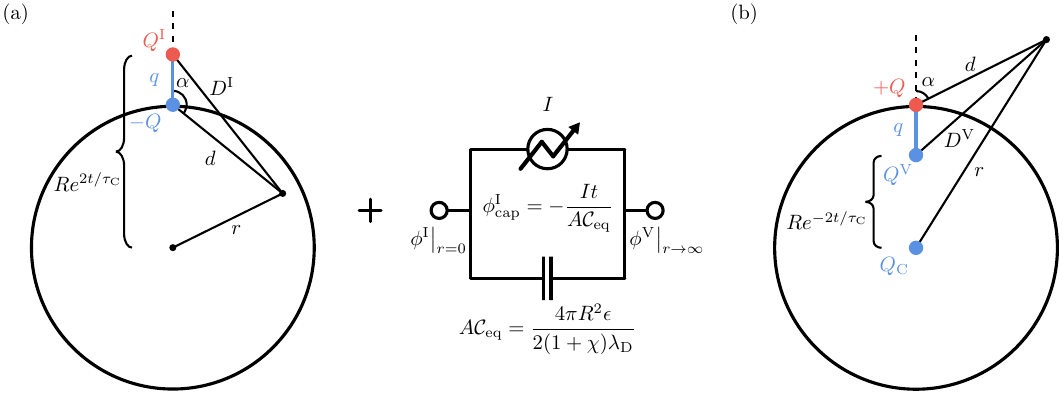}
\caption{
\textbf{Image-charge and equivalent-circuit representation of the bulk potentials.} The bulk potentials {are represented} by a set of effective image charges, as written in Eqs.~\eqref{eq:phi-el-I} and~\eqref{eq:phi-el-V}. The imposed transmembrane current acts as an effective image charge of strength $-Q$ in region $\mathrm I$ and $+Q$ in region $\mathrm V$, where ${Q=2I\tau_{\rm D}}$. (a) In region $\mathrm I$, the pore sink $-Q$ is accompanied by {an} outward-moving image ${Q^{\mathrm I}=Qe^{t/\tau_{\rm C}}}$ at ${r=Re^{2t/\tau_{\rm C}}}$ and a line of charge with linear density ${q=-Q/(2R)}$. An equivalent circuit consisting of the current source and a capacitor {represents} the spatially uniform capacitive contribution to the potential. (b) In region $\mathrm V$, the pore source $+Q$ is accompanied by an inward-moving image ${Q^{\mathrm V}=-Qe^{-t/\tau_{\rm C}}}$ at ${r=Re^{-2t/\tau_{\rm C}}}$, {together with} a line of charge with linear density $q$. {A} central image charge ${Q_{\rm C}=-Q(1-e^{-t/\tau_{\rm C}})^2/2}$ {is also present}. The distances $D^{\mathrm I}$ and $D^{\mathrm V}$ are measured from the moving images to the point of interest, and $\alpha$ is the angle between the polar axis and the vector connecting the source to {that} point.}
    \label{fig:image-charge}
\end{figure*}

The electrostatic contributions to the inner and outer bulk potentials are
\begin{align}
\phi_{\rm el}^{\rm I}
&=
\frac{1}{4\pi\epsilon}
\left[
-\frac{Q}{d}
+\frac{Q^{\rm I}}{D^{\rm I}}
\right.\label{eq:phi-el-I}
\\ 
&\qquad\left.
-\frac{Q}{2R}
\int_{0}^{R(e^{2t/\tau_{\rm C}}-1)}
\frac{\mathrm{d}D'}
{\sqrt{d^2+D'^2-2dD'\cos\alpha}}
\right] ,
\notag
\end{align}
and
\begin{align}
\phi_{\rm el}^{\rm V}
&=
\frac{1}{4\pi\epsilon}
\left[
\frac{Q}{d}
+\frac{Q^{\rm V}}{D^{\rm V}}
+\frac{Q_{\rm C}}{r}
\right.\label{eq:phi-el-V}
\\
&\qquad\left.
-\frac{Q}{2R}
\int_{0}^{R(1-e^{-2t/\tau_{\rm C}})}
\frac{\mathrm{d}D'}
{\sqrt{d^2+D'^2+2dD'\cos\alpha}}
\right] .
\notag
\end{align}
Here, ${Q=2I\tau_{\rm D}}$ is the magnitude of the dominant image charge at the source, $d$ is the distance from the source, and $\alpha$ is the angle between the polar axis and the segment connecting the point of interest to the source. Further, ${Q^{\rm I}=Q \exp(t/\tau_{\rm C})}$ and ${Q^{\rm V}=-Q\exp(-t/\tau_{\rm C})}$ are additional moving image charges that contribute to the inner and outer potentials, respectively, with ${D^{\rm I}}$ and ${D^{\rm V}}$ denoting the corresponding distances from these image charges to the point of interest. The outer potential also contains a contribution from a time-dependent image charge at the center of the vesicle, ${Q_{\rm C}=-Q(1-\exp(-t/\tau_{\rm C}))^2/2}$. The integral term in each expression {represents} a line of charge with linear density ${q = -Q/(2R)}$, extending from the fixed source image charge to the moving image charge. Note that all image charges lie on the boundary or outside {their respective bulk domains, which remain charge-free.}

The capacitive contribution is given by
\begin{equation}
\phi_{\rm cap}^{\rm I}(t)
=
-\frac{(1+\chi)\lambda_{\rm D}}{2\pi R^2\epsilon} \, It
=
-\frac{It}{A\mathcal{C}_{\rm eq}} \ .
\label{eq:cap-contribution}
\end{equation}
where ${A=4\pi R^2}$ is the sphere's surface area. This term is fixed by global charge conservation and describes uniform charging of the membrane and adjacent diffuse charge layers under constant current. It is not associated with a localized image charge and, being spatially uniform, does not contribute to the bulk electric field. Given the form in Eq.~\eqref{eq:cap-contribution}, the capacitive nature of this term is clear: ${It}$ is the total charge separated by the membrane during pumping, while ${A\mathcal{C}_{\rm eq}}$ is the total equivalent capacitance of the vesicle. Figure~\ref{fig:image-charge} shows the locations and magnitudes of the image charges for both the interior and exterior domains, as well as an equivalent circuit that captures the capacitive contribution to the potential.

\begin{figure*}[t]
    \centering
    \includegraphics[width=0.95\linewidth]{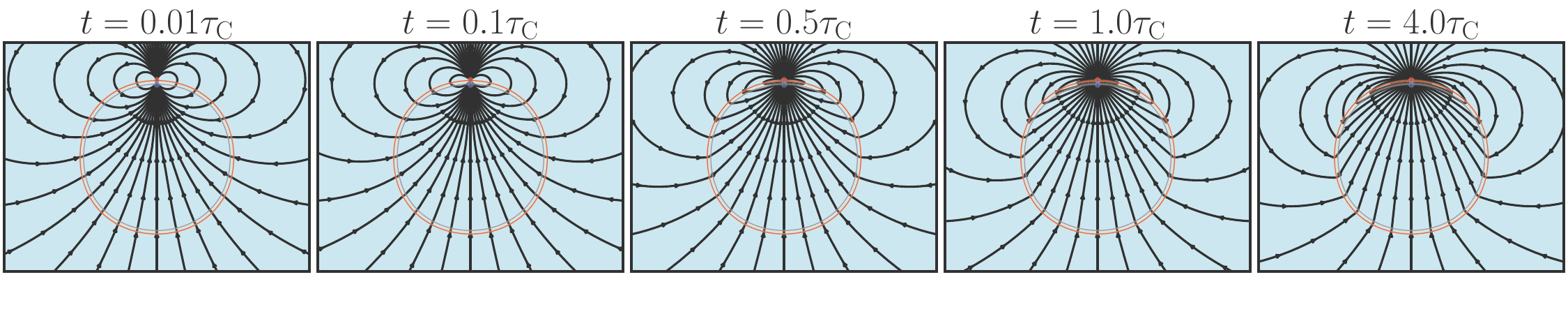}
    \caption{\textbf{Electric-field evolution.} Bulk electric field lines from the analytical solution at ${t = 0.01\,\tau_{\rm C}}$, ${t = 0.1\,\tau_{\rm C}}$, ${t = 0.5\,\tau_{\rm C}}$, ${t = \tau_{\rm C}}$, and ${t = 4\,\tau_{\rm C}}$. At early times ${t\ll \tau_{\rm C}}$, the electric field follows a continuous dipolar behavior on both sides of the membrane. As time advances, capacitive charging of the membrane leads to discontinuities in the field at the membrane surfaces. The interior field becomes monopolar in nature oriented toward the pore and the exterior dipolar field expands. For ${t\gg\tau_{\rm C}}$, the fields reach a steady state, where sustained current increases the overall charge stored on the membrane but no longer changes the electric fields in the bulk domains.}
    \label{fig:elec-fields}
\end{figure*}

{Figure~\ref{fig:elec-fields}} plots the bulk electric fields {corresponding to these potentials} in both domains at selected times. At early times ${t\ll \tau_{\rm C}}$, the {contribution of the} uniform {term} is still small, and the electric fields appear continuous across the membrane. {T}he field lines {resemble} an electric dipole positioned at the source, with dipole moment ${(1+\chi)It\lambda_{\rm D}}$, consistent with the early-time {transmembrane response in} Eq.~\eqref{eq:VM-short-dim}. 
As time {increases}, the electric fields bend at the membrane, reflecting the charging of the diffuse layers and membrane. At later times ${t\gg\tau_{\rm C}}$, the field {inside} the vesicle becomes monopolar in character, with the field lines nearly straight and oriented toward the channel. In the exterior, {the field is also monopolar near the channel,} consistent  with the dominant image charge $Q$ in Eq.~\eqref{eq:phi-el-V}, but remains dipolar far from the channel because the exterior image-charge distribution in Fig.~\ref{fig:image-charge}(b) has no net charge. 

$\\$\textbf{Relaxation after channel closure.}
Because channels and pumps operate for finite durations, we consider the response {after} the current is switched off {at} ${t=t^{\rm o}}$. By linearity, the post-closure response {is a superposition of} the response to {the original} current $I$ {applied at ${t=0}$ and an opposing current $-I$ applied at} ${t=t^{\rm o}}$. {Thus,} ${\phi^\alpha_{\rm rlx}(r,\theta,t) = \phi^\alpha(r,\theta,t) - \phi^\alpha(r,\theta,t - t^{\rm o})}$ for ${t>t^{\rm o}}$ and ${\alpha\in\{{\rm I},{\rm V}\}}$, with an analogous expression for ${V^{\rm M}_{\rm rlx}(d,t)}$.

Physically, channel closure launches an opposing front from the channel at speed $2R/\tau_{\rm C}$. Behind this front, the spatially varying response is canceled and the membrane voltage becomes nearly uniform. Ahead of the front, the voltage remains close to the charging solution of the original current. Both the original and opposing fronts reach the antipode after a time of order $\tau_{\rm C}$ from the onset of their respective currents. For ${t-t^{\rm o}\gtrsim\tau_{\rm C}}$, Eq.~\eqref{eq:VM-long-dim} gives the uniform voltage
${V^{\rm M}_{\rm rlx}=(\chi/(1+\chi))It^{\rm o}/(\pi gR\tau_{\rm C})}$, corresponding to the transported charge being distributed uniformly over the membrane.

Thus, unlike sustained charging, which reaches electrotonic compactness on the longer timescale $\tau_{\rm EC}$, relaxation after closure occurs on the capacitive timescale $\tau_{\rm C}$.

$\\$\textbf{Equivalent circuit model of vesicle charging.}~
The boundary layer problem can also be cast {as} an equivalent circuit for the membrane voltage dynamics. In this picture, each membrane patch stores charge capacitively, while the surrounding electrolyte redistributes charge between patches. Unlike the classical cable equation, however, this coupling is not confined to neighboring patches of the membrane. Instead, current spreads through the surrounding three-dimensional electrolyte, {connecting} distal patches on the membrane.

In Sec.~S4 of the SM, we show that the transmembrane voltage satisfies
\begin{equation}
\mathcal C_{\rm eq}
\frac{\partial V^{\rm M}(\mathbf{x},t)}{\partial t}
=
\mathcal G_{\rm bulk}V^{\rm M}(\mathbf{x},t)
+
\frac{\chi}{\chi+1}\frac{I}{2\pi R^2}
\frac{\delta(\theta)}{\sin\theta} \ ,
\label{eq:cable-nonlocal}
\end{equation}
where the operator $\mathcal{G}_{\rm bulk}$ is
\begin{multline}
\label{eq:cable-kernel}
\mathcal G_{\rm bulk}V^{\rm M}(\mathbf{x},t)
= \\
\mathrm{P.V.}
\int
g_{\rm bulk}(\mathbf{x},\mathbf{y})
\left[
V^{\rm M}(\mathbf{y},t)
-
V^{\rm M}(\mathbf{x},t)
\right]
\mathrm{d}S(\mathbf{y}) \ ,
\end{multline}
with nonlocal conductance kernel $g_{\rm bulk}(\mathbf{x},\mathbf{y})$ given by
\begin{equation}
    g_{\rm bulk}(\mathbf{x},\mathbf{y}) = \frac{g}
{16\pi R^2}
\left(
\frac{4R^2}{|\mathbf{x}-\mathbf{y}|^3}
+
\frac{1}{|\mathbf{x}-\mathbf{y}|}
\right) \ .
\end{equation}
Here, P.V.~denotes the principal value of the integral, which is singular at ${\mathbf{y}=\mathbf{x}}$. Equation~\eqref{eq:cable-nonlocal} {is} a generalized cable equation~\cite{hodgkin1946electrical}. The capacitive current into each membrane patch is balanced by the injected channel current and by nonlocal bulk currents driven by voltage differences between membrane patches. The bulk currents continuously smooth spatial variations in $V^{\rm M}$. The conductance kernel decays algebraically with distance and contains monopolar and dipolar contributions, corresponding to the two {regimes} identified previously~\cite{row2025spatiotemporal}. 

The classical cable equation is recovered when bulk currents are geometrically confined. For example, if the {exterior} electrolyte occupies a thin layer of thickness {$h$}, as in the flat membrane--electrode geometry of Ref.~\cite{fernandes2026electrochemical}, the coupling becomes local, and Eq.~\eqref{eq:cable-nonlocal} reduces to
\begin{multline}
\label{eq:cable-local}
(\mathcal{C}_{\rm eq}+\mathcal{C}_{\rm D})\frac{\partial V^{\rm M}}{\partial t} = g h \nabla^2_{\rm s} V^{\rm M} \\+ \frac{\chi}{\chi+1}\frac{I}{2\pi R^2}\left[\frac{\delta(\theta)}{\sin(\theta)}+\chi+1\right] \ .
\end{multline}
Thus, confinement converts the nonlocal conductance operator of the general case into the local Laplacian of classical cable theory. In both cases, the circuit elements arise directly from electrochemical transport, with capacitance {set by} the membrane and diffuse layers {and} conductance set by the surrounding electrolyte.

\section{Discussion}\label{Sec4}

\noindent\textbf{A single membrane potential as an emergent limit.}~Conventional electrophysiological models often treat a cell as a single electrical compartment with one transmembrane voltage~\cite{hodgkin1946electrical, hodgkin1952quantitative, rall1962theory}. 
Our results show how a spherical vesicle approaches this {limit following the opening of a single channel}. 
Immediately after channel opening, the vesicle cannot be treated as isopotential. Near the source, the transmembrane voltage is governed by the monopolar--dipolar response of a locally flat membrane~\cite{row2025spatiotemporal,fernandes2026electrochemical}. 
Once the outward-moving monopolar--dipolar front has traversed the full surface, the response {becomes} increasingly dominated by spatially uniform capacitive charging of the membrane {and} adjacent diffuse layers. 
Because this contribution is spatially uniform, it does not generate a bulk electric field {inside the sphere}. 
The field is instead determined by the spatially {nonuniform} electrostatic {component}, which {persists} while the channel {is open} and relaxes after closure. Thus, electrotonic compactness is not imposed at the outset but {emerges} as {${t \to \tau_{\rm EC}}$ due to} diffuse-layer charge reorganization {and} bulk conduction on a closed {surface}.
The single-voltage approximation therefore breaks down for $t\lesssim\tau_{\rm EC}$ and near an active channel, where the voltage remains nonuniform.

$\\$\textbf{Relevant timescales.}~
The crossover to the global regime {begins} {on} the capacitive time $\tau_{\rm C}$~\cite{farhadi2025capacitive,zhao2025diffuse}, rather than the diffusion {time} across the vesicle, ${\tau_{\rm R}\sim R^2/D}$. 
The corresponding RC time of a bare electrolyte over the vesicle is ${\tau_{\rm B} \sim \lambda_{\rm D}R/D}$~\cite{bazant2004diffuse}. 
The low-permittivity membrane reduces the effective series capacitance, shortening this timescale by a factor ${(1+\chi)}$, {giving} \({\tau_{\rm C}=\tau_{\rm B}/(1+\chi)}\).
The electrolyte provides the conducting path, while the membrane and 
diffuse layers {provide the} interfacial capacitance. {For micron-scale vesicles and }physiological {parameters}, $\tau_{\rm C}$ is sub-microsecond and may be shorter than typical experimental time resolution.
{M}easurements {may therefore} predominantly {probe the later}, nearly uniform regime. The early spatial structure is not absent, but fast and localized. 
Complete electrotonic compaction {under sustained opening occurs on the longer timescale} ${\tau_{\rm EC}\sim \tau_{\rm C}R/\lambda_{\rm D}}$ {as} the uniform region moves back toward the source. This 
{can take} tens to hundreds of microseconds, {during} which the near-field monopolar regime {persists}. The resulting hierarchy of timescales governing charge reorganization dynamics {is}
\begin{equation}
    \tau_{\rm D} \ll \tau_{\rm C} \ll \tau_{\rm EC} \ll \tau_{\rm R} \  .
\end{equation}

$\\$\textbf{A nonlocal cable equation.}~
The equivalent-circuit formulation
{connects} the spherical response to classical cable theory. Equation~\eqref{eq:cable-nonlocal} has the familiar
structure of capacitive charging balanced by a source current and its lateral
redistribution, but the redistribution operator is nonlocal rather than the
local Laplacian of classical cable theory~\cite{dayan2005theoretical}. Because current spreads through the three-dimensional electrolyte, each membrane patch is coupled to distant patches through a
long-ranged conductance kernel. {The resulting cable equation is therefore nonlocal, with electrical coupling between membrane patches mediated by the surrounding bulk rather than by current confined along the membrane.} Classical cable theory is recovered when the conducting path is
geometrically confined to a thin exterior layer. The nonlocal kernel therefore generalizes the classical cable picture by deriving inter-patch coupling {directly} from diffuse-charge dynamics.

$\\$\textbf{Geometry, caveats, and outlook.}~
More broadly, {our results show that geometry directly shapes} diffuse-charge dynamics. {A} localized transmembrane current {initially produces} spatial voltage variations, but on a closed surface these variations relax toward a global charging mode. This mechanism is absent {for} an unbounded flat membrane, where there is no finite surface over which charge can reorganize to produce a single global voltage.

The sphere provides a simple solvable example in which membrane geometry and closure determine how localized electrochemical perturbations give rise to global voltage dynamics {in a cell}. Several idealizations should be kept in view. We assumed a linearized diffuse-layer response, thin diffuse layers, equal ion diffusivities, no fixed surface charge, and neglected thermal voltage fluctuations~\cite{betancourt2026thermal}. At larger voltages or {for} sustained currents {approaching} $\tau_{\rm R}$, nonlinear effects in the PNP equations, {such as} concentration polarization and field-dependent screening may become important. Future work should extend the nonlocal cable formulation to cylindrical axons and realistic cellular geometries. {This may provide a more general cable theory in which electrical coupling depends explicitly on membrane geometry.}

$\\$\textbf{Acknowledgments.} S.N. acknowledges support from the Gordon and Betty Moore Foundation. K.S. acknowledges support from the McKnight Foundation, the Alfred P. Sloan Foundation, the Philomathia Foundation, and the Camille-Dreyfus Teacher-Scholar Award.
K.K.M. and J.B.F. are supported by Director, Office of Science, Office of Basic Energy Sciences, of the U.S. Department of Energy under contract No.~DE-AC02-05CH11231. J.B.F. is also supported through the Department of Energy Computational Science Graduate Fellowship under Award No.~DE-SC0023112. {All results were originally derived by the authors. Large
language models (Claude and ChatGPT) were used to improve textual clarity in some parts of the manuscript and to {check} intermediate steps in some analytical calculations. Symbolic and numerical calculations were performed using Mathematica and Python. All scientific content, derivations, and conclusions were verified by the authors.} This research used resources of the National Energy Research Scientific Computing Center, a U.S. Department of Energy Office of Science User Facility located at Lawrence Berkeley National Laboratory, under NERSC Award No.~BES-ERCAP0023682.

\bibliography{bibliography}

\end{document}